\documentclass[twocolumn,pre,showpacs]{revtex4}
\usepackage{amssymb}
\usepackage{amsmath}
\usepackage{graphicx}

\def\dbarit {{\mathchar'26\mkern-11mud}}
\input{tcilatex}

\begin{document}

\title{Quantum heat engine with multi-level quantum systems}
\author{H.T. Quan}
\affiliation{Institute of Theoretical Physics, Chinese Academy of Sciences, Beijing,
100080, China}
\author{P. Zhang}
\affiliation{Institute of Theoretical Physics, Chinese Academy of Sciences, Beijing,
100080, China}
\author{C.P. Sun}
\email{ suncp@itp.ac.cn}
\homepage{http://www.itp.ac.cn/~suncp}
\affiliation{Institute of Theoretical Physics, Chinese Academy of Sciences, Beijing,
100080, China}

\begin{abstract}
By reformulating the first law of thermodynamics in the fashion of
quantum-mechanical operators on the parameter manifold, we propose a
universal class of quantum heat engines (QHE) using the multi-level quantum
system as the working substance. We obtain a general expression of work for
the thermodynamic cycle with two thermodynamic adiabatic processes, which
are implied in quantum adiabatic processes. We also classify the conditions
for a 3-level QHE to extract positive work, which is proved to be looser
than that for a 2-level system under certain conditions. As a realistic
illustration, a 3-level atom system with dark state configuration
manipulated by a classical radiation field is used to demonstrate our
central idea.
\end{abstract}

\maketitle

\section{Introduction}

A usual heat engine operates between two heat baths, the temperature
difference between which completely determines the maximum efficiency of the
heat engine. Correspondingly, it is inferred that no power can be extracted
if the two baths have the same temperature. But things become different when
we use quantum matter as the working substance. Quantum effects can
highlight the thermodynamic differences between classical and quantum
working substance of a heat engine. Recently, great efforts \cite{book} has
been devoted to the investigation of the quantum effect of the working
substance. Some exotic phenomena were discovered, most of which concerns the
following three aspects: The first aspect is whether we can improve the
efficiency of the QHE to a level beyond the classical limit. For example, M.
O. Scully et al. \cite{Scully-science,Scully-book,Zubairy-book} proposed a
quantum-electrodynamic heat engine that can exceed the maximum limit by
using the photon gas as working substance. The second aspect is how we can
better the work extraction during a Carnot cycle. T. D. Kieu present a new
type of QHE \cite{Kieu-PRL,Kieu1}, in which the contact time is precisely
controlled (without reaching thermal equilibrium) and the two heat baths are
specifically modified. This QHE can extract more work than the other model
in thermal equilibrium case. The third aspect is about the constraints of
the temperatures of the two heat baths, under which positive work can be
extracted. It is clear that a classical heat engine can extract positive
work when and only when $T_{h}>T_{l}$. Here, $T_{h}$ and $T_{l}$ are the
temperatures of the source and the sink. But for a QHE, some exotic
phenomena may occur. One example is given in Ref. \cite{Scully-science}, in
which positive work can be extracted even from a single heat bath, i.e., $%
T_{h}=T_{l}$; Another example is the simplest 2-level QHE model in Refs. 
\cite{Kosloff1,Kieu1}, in which positive work can be extracted only when $%
T_{h}$ is greater than $T_{l}$ to a certain extent.

We consider the third aspect in detail. In Refs. \cite{Kosloff1,Kieu1}, the
QHE works between the source and sink at temperatures $T_{h}$ and $T_{l}$
respectively. Within a cycle, the level spacing $\Delta $ changes between $%
\Delta ^{h}$ and $\Delta ^{l}$. For such a QHE, the system couples to the
bath for a sufficiently long time until they reach the thermal equilibrium
state. Then positive work can be performed when and only when $%
T_{h}>T_{l}(\Delta ^{h}/\Delta ^{l})$ (condition $I$). This result implies a
broad validity of the second law and shows by how much $T_{h}$ should be
greater\textbf{\ }than $T_{l}$ such that the positive work can be extracted.
This constraints about temperatures is obviously counter-intuitively
different from that of a classical heat engine. Now we wonder whether it is
a universal condition for all multi-level QHE. Actually T. D. Kieu has
considered two special cases, the simple harmonic oscillator and the
infinite square well, in (the appendices of) Ref. \cite{Kieu1}. As to the
two special cases, because all the level spacings change in the same ratio,
the PWC has the same form as that for a 2-level case \cite{Infinite well}.
In this article, we will prove that the PWC for a 3-level QHE can be looser
compared to a 2-level case under our criterion when the level spacings
change properly in the thermodynamic cycle.

This paper is organized as follows: In section $II$, we formulate a quantum
version of the first law of thermodynamics\ for a multi-level quantum
system. In section $III$, we discuss the relationship between the quantum
adiabatic process and the thermodynamic adiabatic process. In section $IV$\
we analyze the quantum thermodynamic cycle of a universal QHE. The obtained
results are more universal than those obtained from the 2-level system \cite%
{Arnaud,He,Kosloff2}. In section $V$, we classify the 3-level QHE according
to the changes of level spacings. We find under certain condition the PWC
for a 3-level QHE can be looser than its counterpart for a 2-level case. A
realistic model---a 3-level atomic system with dark state structure
manipulated by a classical radiation field---is given to demonstrate our
central idea in section $VI$.

\section{Quantum Version of the First Law of thermodynamics}

We consider the QHE with a $N$-level system as its working substance. The
system eigenenergies vary adiabatically with the parameters $\mathbf{R=(}%
R_{1,}R_{2,}...,R_{M,}\mathbf{)}$ in an $M$-dimensional manifold $\mathbf{M}$%
. One can manipulate the parameters to implement the quantum adiabatic
process, which does not excite the transitions among the instantaneous
eigenstates $|n\rangle \equiv $ $|n[\mathbf{R}]\rangle $ of the Hamiltonian $%
\widehat{H}=\widehat{H}[\mathbf{R}]$ with the instantaneous eigenvalues $%
E_{n}$ $\equiv $ $\left\vert E_{n}[\mathbf{R}]\right\rangle $. Usually the
density operator $\widehat{\rho }$ is a function of both the external
parameters $\mathbf{R}$ and the thermodynamic parameter, i.e., the
temperature $T$. So we need to extend the manifold $\mathbf{M}$ to include $%
T $. For example,\ the heat transfer can be also caused by the change of the
temperature. In this sense, we define the differential one-form on the ($%
M+1) $-dimensional \textquotedblleft manifold" $\mathbf{MT}$: $\mathbf{%
\{X=(R,}T\mathbf{)\}}$ by $dF=\nabla _{R}Fd\mathbf{R+}d_{T}F$ for any
function $F$. We emphasize that $d_{T}$ may be a discrete variation $\delta
_{T}$ in the time domain\textbf{.} Then in this sense $\mathbf{MT}$ is no
longer a generic manifold.

During a thermodynamic cycle, work is done by or on the system when the
mechanical parameters $R$\ vary slowly.\textbf{\ }While heat is transferred
when the quantum state or density operator changes as the temperature varies
discretely. At two different instants $t=$ $0$ and $\tau _{2},$ the system
contacts with the source and sink at temperature $T_{h}$ and $T_{l}$
respectively to reach thermodynamic equilibrium. The couplings of the system
to such different heat baths yield the discrete change of the probabilities
distribution in every eigenstate. We consider the infinitesimal variation%
\begin{equation}
dU=Tr(\widehat{\rho }d_{\mathbf{R}}\widehat{H})+Tr(\widehat{H}d\widehat{\rho 
})  \label{1}
\end{equation}%
of the expectation value $U=Tr(\widehat{\rho }\widehat{H})$ of the
Hamiltonian, where we have introduced the density operator of the state $%
\widehat{\rho }$. $dU$ contains two parts corresponding to the changes of
the Hamiltonian and the density operator respectively. Here, the operator
1-form, on the sub-manifold $\mathbf{M}$, can be written as%
\begin{equation}
d_{\mathbf{R}}\widehat{H}=d_{\mathbf{R}}\widehat{Q}+d_{\mathbf{R}}\widehat{W}%
,  \label{2}
\end{equation}%
which is the infinitesimal variation of $\widehat{H}$. Here, the
off-diagonal part%
\begin{eqnarray}
d_{\mathbf{R}}\widehat{Q} &=&\sum_{m\neq n}\langle m|d_{\mathbf{R}}\widehat{H%
}|n\rangle |m\rangle \langle n|  \label{3} \\
&=&\sum_{n}E_{n}d_{\mathbf{R}}(|n\rangle \langle n|)  \notag
\end{eqnarray}%
is the heat operator and the diagonal part%
\begin{eqnarray}
d_{\mathbf{R}}\widehat{W} &=&\sum_{m}\langle m|d_{\mathbf{R}}\widehat{H}%
|m\rangle |m\rangle \langle m|  \label{4} \\
&=&\sum_{m}(d_{\mathbf{R}}E_{m})|m\rangle \langle m|  \notag
\end{eqnarray}%
is the work operator. To derive the above equations we have used the
Feynman-Hellman theorem $\langle m|d\widehat{H}|m\rangle =dE_{m}$ and the
formula $\langle m|d\widehat{H}|n\rangle =$ $(E_{n}-E_{m})\langle
m|d|n\rangle .$

Upon a first glance, the above definitions of the work operator $d_{\mathbf{R%
}}\widehat{W}$ and the heat operator $d_{\mathbf{R}}\widehat{Q}$\ are
reasonable. Intuitively speaking, the work can be done only when the $R$%
-dependent eigenenergy of inner states change. This process is just
described by $d_{\mathbf{R}}\widehat{W}$. The off-diagonal elements of the
infinitesimal variation of $\widehat{H}$ indicates the transitions among the
inner energy levels and thus results in the heat transfer. The term $d_{%
\mathbf{R}}(|n\rangle \langle n|)$ interprets the heat transferred to or
from a quantum system along with the change of the projections to the
instantaneous eigenstate. Physically it means the changing of the occupation
probabilities rather than the $\mathbf{R}$-dependent eigenvalues themselves.
In addition, when the quantum state $\widehat{\rho }$ itself changes due to
the varying of both the control parameters $\mathbf{R}$ and the
thermodynamic parameter $T$, the heat is also transfered. This kind of heat
tansfer is described by the second term on the left hand side of the Eq. (%
\ref{1}).

In comparison with the existing studies about QHE \cite{Kieu1}, we write
down the infinitesimal energy $dU=$ $\dbarit Q+\dbarit W$, where the
infinitesimal heat transferred $\dbarit Q$ and work done $\dbarit W$ are
identified respectively as two path dependent differentials 
\begin{eqnarray}
\dbarit Q &=&Tr(\widehat{H}d\widehat{\rho })+Tr(\widehat{\rho }d_{\mathbf{R}}%
\widehat{Q})=\sum_{m}E_{m}\,dp_{m},  \label{5} \\
\dbarit W &=&Tr(\widehat{\rho }d_{\mathbf{R}}\widehat{W})=\sum_{m}p_{m}%
\,dE_{m}.  \label{6}
\end{eqnarray}%
Here $p_{m}=\langle m|\widehat{\rho }|m\rangle $ are the corresponding
occupation probabilities in the instantaneous states $|m\rangle $.

We would like to emphasize that the above two equations were even obtained
in many previous references \cite{Kieu1,He,Kosloff3}. But here we present a
general derivation, the process of which shows the differences between work
and heat in the view of quantum mechanics. Substantially, we give the
microscopic definitions of work and heat based on the non-adiabatic
transitions among the instantaneous eigenstates of the time-dependent
Hamiltonian.

\section{From quantum adiabatic evolution to thermodynamic adiabatic process}

The term \textquotedblleft adiabatic process" is usually used in both
quantum mechanics and thermodynamics, and seemingly has different meanings
for these two cases. Actually, a quantum adiabatic process implies a
thermodynamic adiabatic process. But not all thermodynamic adiabatic
processes is caused by quantum adiabatic processes \cite{Kieu1,yukawa}. This
understanding is crucial for the following analysis about the thermodynamic
cycle of the QHE.

In quantum mechanics, the adiabatic process is described by the time
evolution of a quantum system with slowly changing parameters. When these
parameters vary slow enough, the transitions among the instantaneous
eigenstates $|n[\mathbf{R}]\rangle \equiv $ $|n(t)\rangle $ are forbidden,
i.e., the system will keep in the $n$-th instantaneous eigenstate $%
|n(t)\rangle $ if the system is initially in the eigenstate $|n(0)\rangle $
at time $t=0$. Generally speaking, starting with the initial state $|\psi
(0)\rangle =\sum\limits_{n}c_{n}|n(0)\rangle $, the system will evolve into%
\begin{equation}
|\psi (t)\rangle =\sum\limits_{n}c_{n}\exp [-i\int_{0}^{t}E_{n}(t^{\prime
})dt+i\gamma _{n}]|n(t)\rangle ,  \label{7}
\end{equation}%
where%
\begin{equation*}
\gamma _{n}=\int_{0}^{t}\left\langle n(t^{\prime })\right\vert \frac{d}{%
dt^{\prime }}\left\vert n(t^{\prime })\right\rangle dt^{\prime }
\end{equation*}%
is the so-called Berry's phase. This conclusion implies that the occupation
probability $\left\vert \left\langle n(t)\right. \left\vert \psi
(t)\right\rangle \right\vert ^{2}$ in an instantaneous eigenstate $%
|n(t)\rangle $ is adiabatically invariant. This result is also valid for the
initial mixed state, e.g., the thermal equilibrium state%
\begin{equation}
\widehat{\rho }(0)=\sum_{n}\rho _{nn}|n(0)\rangle \langle n(0)|,  \label{8}
\end{equation}%
where $\rho _{nn}$ are Gibbs probability distributions.

We illustrate a quantum adiabatic process and its corresponding
thermodynamic adiabatic process in Fig. 1. An amount of ideal gas is
constrained in a cylinder. The piston moves so slow that the gas is always
kept in thermal equilibrium state. Usually no heat is gained or lost in a
thermodynamic adiabatic process. Thus any isotropic process implies its
adiabatic property. But all adiabatic process are not isentropic. For
example, an adiabatic-free expansion is not isentropic. Now we prove a
quantum adiabatic process illustrated in Fig. 1b microscopically leads to an
isentropic process and thus a thermodynamic adiabatic process.

\ 
\begin{figure}[h]
\begin{center}
\includegraphics[bb=53 333 554 788, width=8 cm, clip]{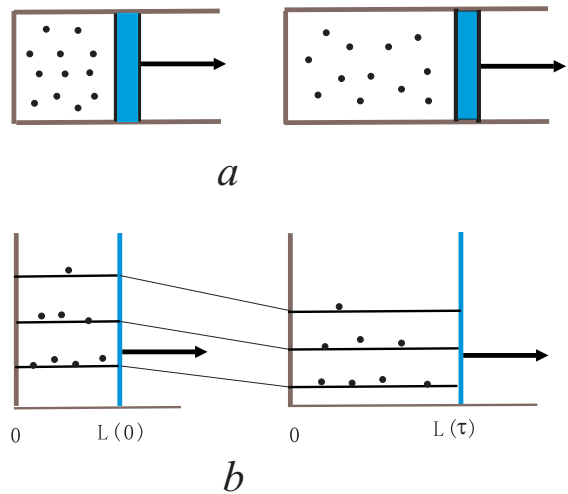}
\end{center}
\caption{(color on line) Illustration of classical adiabatic process ($a$)
and its quantum-mechanical counterpart ($b$). At time $t=0$, the piston
(wall) locates at $L(0)$, and after a long time $t=\protect\tau $, the
piston (wall) moves slowly to $L(\protect\tau )$. No change happens in the
energy level populations during the adiabatic process.}
\label{fig.1}
\end{figure}
\bigskip\ \ \ 

In quantum mechanics, we describe the motion of these structureless gas
atoms or molecules with an infinite potential well with one moving boundary
(Fig. 1b). The piston or boundary moves so slow that the quantum adiabatic
condition are satisfied \cite{adiabatic condition}. Now we consider the
microscopic definition of the thermodynamic entropy%
\begin{equation}
S=-kTr\left( \widehat{\rho }[\mathbf{R}]\ln \widehat{\rho }[\mathbf{R}%
]\right) .  \label{9}
\end{equation}%
The variation of the entropy due to the changes of the mechanical parameters 
$\mathbf{R}$ can be calculated as%
\begin{equation}
d_{\mathbf{R}}S[\mathbf{R}]=-k\sum\limits_{n}\left( \rho _{nn}\ln \rho
_{nn}+1\right) d_{\mathbf{R}}\rho _{nn}.  \label{10}
\end{equation}%
According to the quantum adiabatic theorem, the particle distributions in
the instantaneous energy levesls are invariant during a quantum adiabatic
process. Then, from Eq. (\ref{10}), the entropy keeps unchanged. Therefore,
it is concluded that the quantum adiabatic process of the microscopic
particles just results in the thermodynamic adiabatic process of the
macroscopic system.

The above arguments shows that the quantum adiabatic process in quantum
mechanics can result in a thermodynamic adiabatic process. But we have to
point out that not all thermodynamic adiabatic process are caused by the
quantum mechanical adiabatic process. For example, in the thermodynamic
adiabatic process of Fig. 1$a$ a single molecule may experience nonadiabatic
transitions due to its interaction with other molecules.

\section{Positive work done by multilevel QHE in a thermodynamic cycle}

\emph{\ }Like the classical heat engine, a universal QHE also bases on a
thermodynamic cycle. Mathematically, it can be understood as a close path $l$%
: $\{\mathbf{X}=(\mathbf{R},T)|\mathbf{X(0)=X(\tau }_{4}\mathbf{)}\}$ on the
($M+1)$-dimensional \textquotedblleft manifold" $\mathbf{MT.}$ The
efficiency of a universal QHE is $\eta _{q}=\Delta W/Q$ \ corresponding to
the ratio of the work done to the heat absorbed during a cycle,%
\begin{eqnarray}
\Delta W &=&\oint \dbarit W  \label{11} \\
Q &=&\int_{\mathbf{0}}^{\tau _{1}}\dbarit Q,  \label{12}
\end{eqnarray}%
where $\tau _{1}$ is the ending point of \emph{Step 1}. A typical
four-stroke QHE, consists of two quantum adiabatic and two isothermal
processes, is a quantum analogue of the classical Otto engine, as
illustrated in Fig. 2:

\ 
\begin{figure}[h]
\begin{center}
\includegraphics[bb=37 185 552 676, width=7 cm, clip]{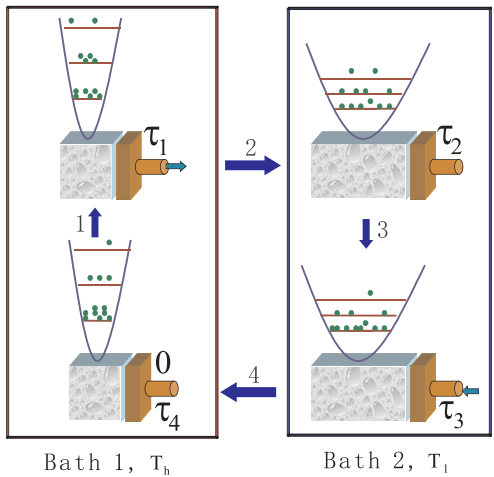}
\end{center}
\caption{(color on line) The graphic sketch of a universal QHE model. From
time $t=0$ to t=$\protect\tau _{1}$, the system absorbs heat from the source
(bath 1). From time $t=\protect\tau _{1}$ to t=$\protect\tau _{2}$, work is
done by the system when the piston is pushed by the working substance. Step
3 (from time $t=\protect\tau _{2}$ to t=$\protect\tau _{3}$) and step 4(from
time $t=\protect\tau _{3}$ to t=$\protect\tau _{4}$) are reversed processes
of step 1 and step 2 with modification.}
\label{fig.2}
\end{figure}

\bigskip

\emph{Steps 1 }and \emph{Step 3 }are two isothermal processes. During \emph{%
Step 1},\emph{\ }i.e., from time $t=0$ to $t=\tau _{1}$, the system couples
to the source (bath 1) at temperature $T_{h}$ and the level spacings keep
unchanged. After absorbing energy from the heat source, the system reaches
thermodynamic equilibrium state, which can be described by the density
operator $\widehat{\rho }(\tau _{1})$. The heat absorbed during \emph{Step 1}
is 
\begin{equation}
Q=\int_{\mathbf{0}}^{\tau _{1}}\sum_{m}E_{m}\langle m|\delta _{T}\widehat{%
\rho }|m\rangle .  \label{13}
\end{equation}%
\emph{Step 3 }is almost an inverse process of \emph{Step 1}. From time $\tau
_{2}$\ to $\tau _{3}$, the system is brought to couple to the sink at
temperature $T_{l}$. The density matrix changes into $\widehat{\rho }(\tau
_{3})$ after releasing energy into the bath. In the above operations, the
absorption and the release of energy during \emph{Steps 1 }and \emph{Step 3}
happen in quantum ways. Namely, they can only happen probabilistically. The
corresponding probability depends on the details of the interactions and
some intrinsic properties, e.g., the temperatures of the heat baths.

\emph{Steps 2 }and \emph{Step 4 }are two adiabatic processes. During \emph{%
Steps 2} the QHE performs positive work when the energy spacings decrease.
Meanwhile the parameters $\mathbf{R}$ adiabatically change from $\mathbf{R}%
(\tau _{1})=\mathbf{R}(0)$ to $\mathbf{R}(\tau _{2})$. But the atomic
probability distribution remains unchanged in this process. \emph{Step 4 }is
almost an inverse process of \emph{step 2}, during which the system is
removed from the sink and its energy gaps increase as an amount of work is
done on the system. The net work done by the system during a cycle is%
\begin{equation}
\Delta W=\left( \int_{\tau _{1}}^{\tau _{2}}+\int_{\tau _{3}}^{\tau
_{4}}\right) \sum_{m}p_{m}dE_{m}.  \label{14}
\end{equation}%
The above results can give the known results in \cite{Kieu1,He,Kosloff3} for
a 2-level system.

We assume the thermal equilibrium Gibbs distributions for the heat bath. The
system will eventually reach Gibbs distribtution%
\begin{equation}
\widehat{\rho }(\tau )=Z^{-1}e^{-\beta _{s}\widehat{H}}=\sum_{m}p_{m}^{s}|m%
\rangle \langle m|  \label{15}
\end{equation}%
after coupling to the heat bath for a time much longer than the relaxation
time $\gamma ^{-1}$ of the considered system. The occupation probability in
the instantaneous eigenstate $|m\rangle $ is%
\begin{equation}
p_{m}^{s}(t)=\frac{e^{-\beta _{s}E_{m}(T_{s})}}{Z^{s}},  \label{16}
\end{equation}%
which depends on the spectral structure and the temperature $T_{s}(s=h,l)$
of the relevant heat baths. Here $\beta _{s}=1/(KT_{s})$, and $K$ is the
Boltzmann constant.$\ $The partition function is defined by 
\begin{equation}
Z^{s}=\sum_{m}e^{-\beta _{s}E_{m}(T_{s})}.  \label{17}
\end{equation}%
For the cyclic nature of QHE, the energy level spacings return at different
instants. Then the net work done during a cycle can be calculated as%
\begin{equation}
\Delta W=\sum_{m}\left[ p_{m}^{h}(\tau _{1})-p_{m}^{l}(\tau _{3})\right]
\Delta E_{m},  \label{18}
\end{equation}%
where 
\begin{equation}
\Delta E_{m}=E_{m}(T_{h})-E_{m}(T_{l}).  \label{19}
\end{equation}%
Therefore, the PWC can be explicitly expressed as 
\begin{equation}
\Delta W=\sum_{m}\left[ p_{m}^{h}(\tau _{1})-p_{m}^{l}(\tau _{3})\right]
\Delta E_{m}>0.  \label{20}
\end{equation}

\section{Classification of positive work cycles for 3-level QHE}

It is also known from Refs. \cite{Kieu1,Kosloff3} that the PWC for the
2-level QHE working between the source and sink at temperature $T_{h}$ and $%
T_{l}$ can be written as%
\begin{equation}
T_{h}>T_{l}\frac{\Delta ^{h}}{\Delta ^{l}}.  \label{21}
\end{equation}%
Only when this PWC is satisfied can the positive work be extracted. This
condition is counter-intuitively different from that $T_{h}>T_{l}$ for a
classical heat engine. Now, we naturally ask a question: Is this condition (%
\ref{21}) universal for a multi-level QHE? In this section we will prove
that, if the energy levels change properly, the PWC for a 3-level system can
be looser than that for a 2-level system (\ref{21}).

Our model considered here is a 3-level system with the adjustable level
spacings $\Delta _{1}^{s}$ and $\Delta _{2}^{s}$ $(s=h,l)$, as illustrated
in Fig. 3. Here, we denote the ground state, the first excited state and the
second excited state with subscripts $0$, $1$ and $2$ respectively. We also
introduce the dependent parameters $\Delta ^{s}=\Delta _{1}^{s}+\Delta
_{2}^{s},(s=h,l)$. For this kind of 3-level systems, if the PWC (\ref{20})
can be reduced to 
\begin{equation}
\left\{ 
\begin{array}{c}
T_{h}>T_{l}\left( \Delta ^{h}/\Delta ^{l}\right) \theta \\ 
T_{h}>T_{l}\left( \Delta _{1}^{h}/\Delta _{1}^{l}\right) \theta ^{\prime }%
\end{array}%
\right. ,  \label{22}
\end{equation}%
meanwhile both $\Delta ^{l}/\Delta ^{h}<\theta <1$ and $\Delta
_{1}^{l}/\Delta _{1}^{h}<\theta ^{\prime }<1$ are satisfied, then \emph{we
say the PWC is looser than that for a 2-level case} since the PWCs for both
the two substructures are looser than that for a 2-level system (\ref{21}).
The two substructures are formed by combining level $0$ with level $1$ and
level $2$ respectively. We will prove in the following there indeed exist
such cases for a 3-level system. This means that the PWC for a 3-level
system can be improved in comparison with that for a 2-level case when the
levels change properly.

\begin{figure}[tbp]
\includegraphics[bb=47 563 383 751, width=8 cm, clip]{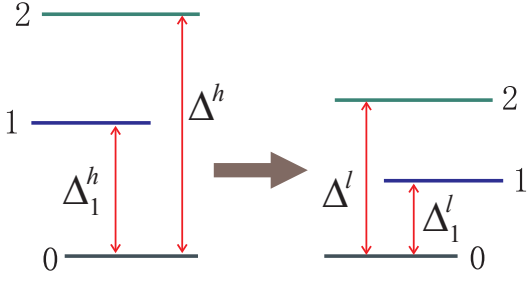}
\caption{(color on line) A sketch of an adiabatic evolution of the level
spacings of a 3-level system during the step 2. $\Delta _{i}^{h}$ and $%
\Delta _{i}^{l}(i=1,2)$ are the two level spacings when coupling to the
source and the sink. $\Delta ^{h}(\Delta ^{l})$ is the maximum level
spacing, i.e., $\Delta ^{s}=\Delta _{1}^{s}+\Delta _{2}^{s},(s=h,l)$.}
\end{figure}

For the 3-level case we rewrite the expression (\ref{18}) of $\Delta W$ in a
more explicit form 
\begin{equation}
\Delta W=\frac{G(\Delta ,\Delta _{1})+G(\Delta _{1},\Delta )}{%
Z^{h}(T_{h})Z^{l}(T_{l})},  \label{23}
\end{equation}%
where $Z_{s}(T_{s})$ is the partition function for the three level case.%
\begin{eqnarray}
G(\Delta ,\Delta _{1}) &=&\Theta (\Delta ,\Delta _{1})(\Delta ^{h}-\Delta
^{l}),  \label{24} \\
\Theta (\Delta ,\Delta _{1}) &=&e^{-\beta _{h}\Delta ^{h}}-e^{-\beta
_{l}\Delta ^{l}} \\
&&+e^{-\beta _{h}\Delta ^{h}-\beta _{l}\Delta _{1}^{l}}-e^{-\beta _{h}\Delta
_{1}^{h}-\beta _{l}\Delta ^{l}}.  \notag
\end{eqnarray}%
Obviously $\Delta W$ is completely determined by $T_{s},$ $\Delta _{1}^{s}$
and $\Delta _{2}^{s}(s=h,l)$, which is independent of the details of the
changing of the level spacings. Thus, we only care about the initial level
spacing $\Delta _{i}^{h}$ and the final level spacing $\Delta _{i}^{l}$ in
considering the thermodynamic cycle.

However, at arbitrary finite temperature, the PWC (\ref{20}) is too
complicated to be understood directly, so we switch to considering its high
temperature limit. In this limit the PWC (\ref{20}) can be simplified as%
\begin{equation}
L\left[ F(\xi ,\eta )-\frac{T_{l}}{T_{h}}\frac{\Delta ^{h}}{\Delta ^{l}}%
F(\xi ,\lambda )\right] \left( \Delta _{1}^{h}-\Delta _{1}^{l}\right) >0,
\label{25}
\end{equation}%
where we have introduced 
\begin{eqnarray}
F(\xi ,\eta ) &=&\left[ \left( 2\xi -1\right) +\left( 2-\xi \right) \eta %
\right] ,  \notag  \label{25.5} \\
L &=&\frac{\Delta ^{l}}{kT_{l}Z^{h}(T_{h})Z^{l}(T_{l})}
\end{eqnarray}%
in terms of an independent set of parameters%
\begin{eqnarray}
\xi &\equiv &1+\frac{\Delta _{2}^{h}-\Delta _{2}^{l}}{\Delta _{1}^{h}-\Delta
_{1}^{l}};  \label{26} \\
\text{ }\eta &\equiv &\frac{\Delta _{1}^{l}}{\Delta ^{l}};\text{ }\lambda
\equiv \frac{\Delta _{1}^{h}}{\Delta ^{h}}.  \notag
\end{eqnarray}%
According to the requirement of the quantum adiabatic evolution, the level
spacings $\Delta _{1}^{s}$\ and $\Delta _{2}^{s}$\ are always kept positive
during the two quantum adiabatic process. Therefore, the parameters $\eta $
and $\lambda $ range in the interval $(0,1)$, i.e., $0<\eta ,$ $\lambda <1$.

In principle, we can classify the multi-level QHE according to the changes
of the level spacings in the thermodynamic cycles. As to the 3-level QHE,
there are altogether four sorts of operations corresponding to the following
four cases:%
\begin{equation}
\left\{ 
\begin{array}{c}
(I).\Delta _{1}^{h}-\Delta _{1}^{l}>0,\text{ }\Delta _{2}^{h}-\Delta
_{2}^{l}>0 \\ 
(II).\Delta _{1}^{h}-\Delta _{1}^{l}>0,\text{ }\Delta _{2}^{h}-\Delta
_{2}^{l}<0 \\ 
(III).\Delta _{1}^{h}-\Delta _{1}^{l}<0,\text{ }\Delta _{2}^{h}-\Delta
_{2}^{l}>0 \\ 
(IV).\Delta _{1}^{h}-\Delta _{1}^{l}<0,\text{ }\Delta _{2}^{h}-\Delta
_{2}^{l}<0%
\end{array}%
\right. ,  \label{27}
\end{equation}%
as schematized in Fig. 4.

\begin{figure}[tbp]
\includegraphics[bb=7 241 573 708, width=8 cm, clip]{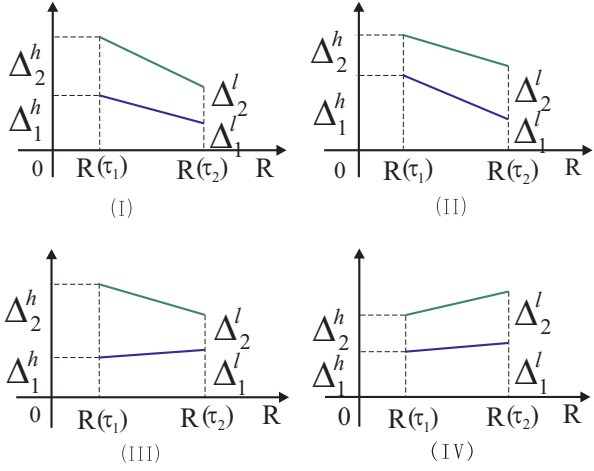}
\caption{(color online) Sketch map of four sorts of 3-level QHE. There are
altogether four kinds of 3-level QHE, as illustrated above, according to the
changes of the level spacings during step 2. Here (I), (II), (III) and (IV)
are corresponding to case (I) to case (IV) in Eq. (29) respectively.}
\end{figure}

Before analyzing the four cases in detail, we intuitively consider the
physical mechanism that a 3-level QHE may better the work done. Namely, the
PWC can be relaxed compared to that for a 2-level case. For instance, if the
lower level spacing does not change during \emph{step 2} (see Fig. 5 $a$), $%
\Delta _{1}^{h}=\Delta _{1}^{l}$ (dashed line), the 3-level QHE reduced to a
2-level case, then the PWC cannot be bettered. However, when it comes to $%
\Delta _{1}^{h}>\Delta _{1}^{l}$\ (solid line), the lowing of energy level $%
1 $ produces extra work to better the PWC. On the contrary, if the level
spacings changes as that in Fig. 5 $b$, $\Delta _{1}^{h}<\Delta _{1}^{l}$ (
solid line), the raising of the energy level $1$ makes the PWC become worse.
These two cases are corresponding to \textbf{Case (I)}\ and \textbf{Case
(III)}\ in Fig. 4. We expect that the 3-level QHE of \textbf{Case (I)}\ can
better the work extraction but \textbf{Case (III) }can not. We will try to
prove this result and determine the physical parameters that can better the
work extraction in the following detailed analysis. Firstly we consider 
\textbf{Case (I)}.

\begin{figure}[tbp]
\includegraphics[bb=29 465 587 722, width=8 cm, clip]{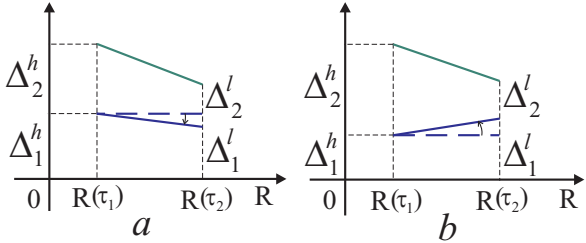}
\caption{(color online) A 3-level QHE built up by adding a third level to
the 2-level QHE. Intuitively speaking, it can make the PWC looser (better
the work extraction) when the new level helps to produce positive work ($a$%
), but makes the PWC worse when it cost energy ($b$).}
\end{figure}

\textbf{Case (I)}: Physically, $\Delta _{1}^{h}-\Delta _{1}^{l}>0$ and $%
\Delta _{2}^{h}-\Delta _{2}^{l}>0$ mean both of the two level spacings
decrease adiabatically during \emph{step 2}. Since $\Delta _{1}^{h}-\Delta
_{1}^{l}>0$, the PWC (\ref{25}) can be simplified as%
\begin{equation}
F(\xi ,\eta )T_{h}>T_{l}F(\xi ,\lambda )\frac{\Delta ^{h}}{\Delta ^{l}}.
\label{28}
\end{equation}%
By noticing that $\xi >1$, $F(\xi ,\eta )>1$ and $F(\xi ,\lambda )>1$ in 
\textbf{Case (I)}, the above PWC (\ref{28}) can be further simplified as 
\begin{equation}
T_{h}>T_{l}\frac{\Delta ^{h}}{\Delta ^{l}}\theta .  \label{29}
\end{equation}%
Here, the dimensionless parameter $\theta $ is defined by 
\begin{equation}
\theta =\frac{F(\xi ,\lambda )}{F(\xi ,\eta )}.  \label{30}
\end{equation}

Aiming to construct a 3-level QHE with a relaxed constraints (\ref{22}) on
temperatures, we need to analyze in what conditions $\theta <1$ is
satisfied. Obviously, as to $\theta <1$, i.e., $\left( 2-\xi \right) \lambda
<\left( 2-\xi \right) \eta $, there are two solutions. Solution $I$:%
\begin{equation}
\xi >2\ and\ \lambda >\eta ,  \label{31}
\end{equation}%
\ and Solution $II$:%
\begin{equation}
1<\xi <2\ and\ \lambda <\eta .  \label{32}
\end{equation}%
For simplicity, we define a set of independent physical parameters in terms
of the ratios of three level spacings to $\Delta _{1}^{h}$,%
\begin{equation}
r_{1}^{l}=\frac{\Delta _{1}^{l}}{\Delta _{1}^{h}},r_{2}^{l}=\frac{\Delta
_{2}^{l}}{\Delta _{1}^{h}},r_{2}^{h}=\frac{\Delta _{2}^{h}}{\Delta _{1}^{h}}.
\label{33}
\end{equation}%
Then it is easy to see that%
\begin{equation}
r_{1}^{l}<1,r_{2}^{h}>r_{2}^{l}  \label{33.5}
\end{equation}%
are implied in our presupposition of \textbf{Case (I)}.

Based on the above analysis, Solution $I$ can be explicitly obtained to be%
\begin{equation}
\left\{ 
\begin{array}{c}
r_{2}^{h}+r_{1}^{l}-r_{2}^{l}>1 \\ 
r_{2}^{l}>r_{1}^{l}r_{2}^{h}%
\end{array}%
\right. .  \label{34}
\end{equation}%
We combine Eqs. (\ref{33.5}) and (\ref{34}) and plot them in a 3-dimensional
figure in coordinates of $r_{1}^{l}$, $r_{2}^{l}$ and $r_{2}^{h}$. The
solution is a 3-dimensional domain enveloped by a curved surface, $%
r_{2}^{l}=r_{1}^{l}r_{2}^{h}$ (denoted by $A$) and two planes $%
r_{2}^{h}=1-r_{1}^{l}+r_{2}^{l}$ (denoted by $B$) and $r_{1}^{l}=0$. Any
representative point in this domain can determine a 3-level QHE, which can
extract positive work under the conditions $T_{h}>T_{l}\left( \Delta
^{h}/\Delta ^{l}\right) \theta $ and $\Delta ^{l}/\Delta ^{h}<\theta <1$.

We further consider whether Solution $I$ is consistent with the second
inequality of Eq. (\ref{22}). Fortunately we can easily find such $\theta
^{\prime }$ smaller than unity for $\Delta _{1}^{h}/\Delta _{1}^{l}>\Delta
^{h}/\Delta ^{l}$ in Solution $I$. Thus, according to the the definition (%
\ref{22}), it is proved that the PWC for 3-level system can be looser than
that for a 2-level case if the level spacings change properly.

\begin{figure}[tbp]
\includegraphics[bb=230 437 593 749, width=8 cm, clip]{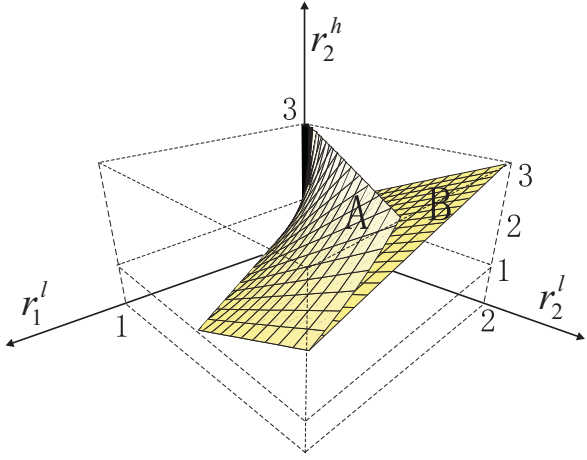}
\caption{(color online) Diagram of solution I of case I. For a given $\Delta
_{1}^{h}$, when $\Delta _{1}^{l},\Delta _{2}^{l}$ and $\Delta _{2}^{h}$ are
properly chosen such that the point $(r_{1}^{l},r_{2}^{l},r_{2}^{h})$ is in
the enveloped range, positive work can be extracted under a looser
condition. }
\end{figure}

Similarly we rewrite Solution $II$ (\ref{32}) in terms of $r_{1}^{l}$, $%
r_{2}^{l}$ and $r_{2}^{h}$,%
\begin{equation}
\left\{ 
\begin{array}{c}
r_{2}^{h}+r_{1}^{l}-r_{2}^{l}<1 \\ 
r_{2}^{l}<r_{1}^{l}r_{2}^{h}%
\end{array}%
\right. .  \label{35}
\end{equation}%
Together with (\ref{33.5}), we illustrate the result in Fig. 7. The physical
meaning is the same as that of Solution $I$. But Solution $II$ is not an
ideal solution like Solution $I$, for we cannot find such solution that
satisfy the second inequality.

\begin{figure}[tbp]
\includegraphics[bb=266 455 566 680, width=8 cm, clip]{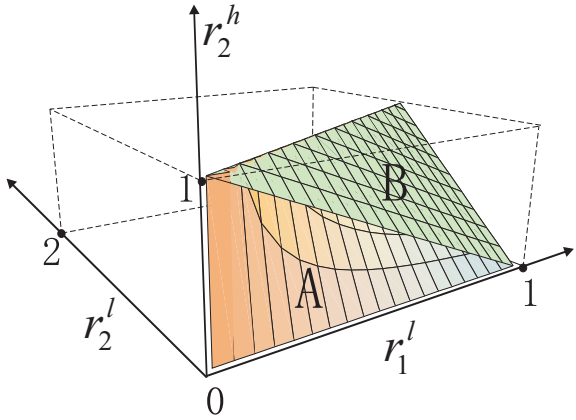}
\caption{(color online) Diagram of solution II of case I. The representative
points in the enclosed range are the results of inequalities(\protect\ref{35}%
), and the physical meaning is similar to that of Fig. 6.}
\end{figure}

In order to make our result clear, we illustrate the above results in
comparison with\textbf{\ }that for a classical heat engine and a 2-level QHE
in Fig. 8 and Fig. 9. As to a classical heat engine, e.g., a Carnot engine,
the PWC is $T_{h}>T_{l}$, and the net work $\Delta W$ is a linear function
of $T_{h}$ for a given $T_{l}$. But for a 2-level QHE, the PWC is Eq. (\ref%
{21}), and the net work is not a linear function of $T_{h}$, as illustrated
in Fig. 8.

\begin{figure}[tbp]
\includegraphics[bb=161 191 425 494, width=6 cm, clip]{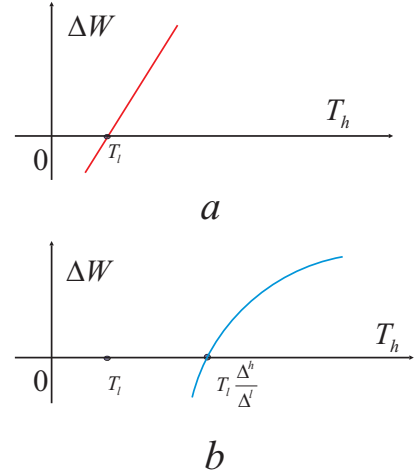}
\caption{(color online) Sketch map of the net work $\Delta W$ as a function
of $T_{h}$ for a given $T_{l}$. Part $a$ illustrates a classical Carnot
engine, while part $b$ illustrates a 2-level QHE.}
\end{figure}
As to Solution $I$ of case $I$ of the 3-level QHE, the PWC is looser than
that for a 2-level case, because\textbf{\ }the PWC for both the two
substructures are looser than that for the 2-level system, as illustrated in
Fig. 9$a$. However, as to Solution $II$, the PWC is not looser than that for
a 2-level case, for the PWC for one substructure is looser than that for the
2-level system, but the other is not, as illustrated in Fig. 9$b$.

\begin{figure}[tbp]
\includegraphics[bb=136 364 446 746, width=6 cm, clip]{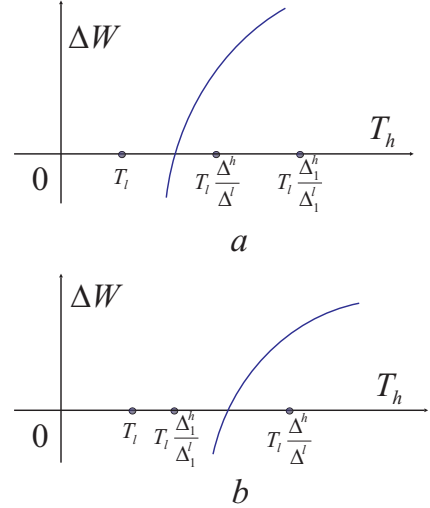}
\caption{(color on line) Similar to Fig. 8. Part $a$ illustrates Solution $I$
of Case $I$ of the 3-level QHE, and Part $b$ illustrates Solution $II$ of
Case $I$ of the 3-level QHE.}
\end{figure}

\textbf{Case (II):} In this case $\Delta _{1}^{h}-\Delta _{1}^{l}>0$ and $%
\Delta _{2}^{h}-\Delta _{2}^{l}<0.$ Physically, this means that the upper
level spacing increases while the lower level spacing decreases during \emph{%
step 2}. We will prove in the following in this case there exist no such
solution about $\Delta _{i}^{h}$ and $\Delta _{i}^{l}$ that can better the
work extraction compared to a 2-level case.

To prove this conclusion we need to distinguish the following 4 situations:%
\begin{eqnarray}
\mathbf{a} &:&F(\xi ,\eta )>0,F(\xi ,\lambda )>0,  \label{35.1} \\
\mathbf{b} &:&F(\xi ,\eta )>0,F(\xi ,\lambda )<0,  \notag \\
\mathbf{c} &:&F(\xi ,\eta )<0,F(\xi ,\lambda )<0,  \notag \\
\mathbf{d} &:&F(\xi ,\eta )<0,F(\xi ,\lambda )>0.  \notag
\end{eqnarray}

\textbf{a}: The PWC (\ref{25})\ can \ be reduced to Eq. (\ref{29}). Similar
to above analysis, if we want to find a looser PWC, $\theta $ should be
smaller than unity, i.e.,$\ \theta <1$. By noticing $\xi <1$ in \textbf{Case
(II)}, $\theta <1$ can be further simplified as $\lambda <\eta $, i.e., 
\begin{equation}
\frac{\Delta _{1}^{l}}{\Delta _{1}^{h}}>\frac{\Delta _{2}^{l}}{\Delta
_{2}^{h}}.  \label{35.3}
\end{equation}%
However, this inequality is incompatible with the two constraints of \textbf{%
Case (II):} $\Delta _{1}^{h}-\Delta _{1}^{l}>0$ and $\Delta _{2}^{h}-\Delta
_{2}^{l}<0$. Thus, in this situation, the PWC is no looser than that for a
2-level QHE for $\theta >1$.

\textbf{b}: The two inequalities $F(\xi ,\eta )>0$ and $F(\xi ,\lambda )<0$
can be reduced to%
\begin{equation}
\frac{1-2\eta }{2-\eta }<\xi <\frac{1-2\lambda }{2-\lambda }.  \label{35.4}
\end{equation}%
Thus we can obtain the inequality $\eta >\lambda $, i.e., 
\begin{equation*}
\frac{\Delta _{1}^{l}}{\Delta _{1}^{h}}>\frac{\Delta _{2}^{l}}{\Delta
_{2}^{h}}.
\end{equation*}%
Similarly, this inequality is incompatible with $\Delta _{1}^{h}-\Delta
_{1}^{l}>0$ and $\Delta _{2}^{h}-\Delta _{2}^{l}<0$. Thus, in \textbf{Case
(II)} $F(\xi ,\eta )>0$ and $F(\xi ,\lambda )<0$ can not be satisfied
simultaneously. Therefore, there does not exist the PWC looser than that for
a 2-level QHE in this situation.

\textbf{c}: The PWC can now be reduced to be%
\begin{equation}
T_{h}<T_{l}\frac{\Delta ^{h}}{\Delta ^{l}}\theta .  \label{35.6}
\end{equation}%
By noting that we have assumed beforehand $T_{h}>T_{l}$. Only when $(\Delta
^{h}/\Delta ^{l})\theta >1$, the QHE is probable to extract positive work.
From Eq. (\ref{35.1}), $F(\xi ,\eta )<0$ is always satisfied in condition 
\textbf{c}. Then $(\Delta ^{h}/\Delta ^{l})\theta >1$ can be reduced to%
\begin{equation}
\xi ^{2}-\xi +1=(\xi -\frac{1}{2})^{2}+\frac{3}{4}<0.  \label{35.8}
\end{equation}%
Obviously this inequality has no solution to the real parameter $\xi $.
Therefore the positive work can not be extracted in this situation.

\textbf{d}: The PWC can now be reduced to $T_{h}<T_{l}(\Delta ^{h}/\Delta
^{l})\theta .$ The left hand side in the inequality is positive while the
right hand side is negative (for $\theta <0$). It is obvious that, there
does not exist a PWC looser than that for a 2-level QHE in this situation,
either.

In summary, we cannot find a PWC looser than that for a 2-level case in 
\textbf{Case (II)}. After similar analysis about \textbf{Case (III) }and 
\textbf{Case (IV)}, we find there is no desired solution in \textbf{Case
(III) }and \textbf{Case (IV)} either. In conclusion, only in \textbf{Case (I)%
} can we find such desired solutions that the 3-level QHE can better the
work extraction compared to the 2-level case. This conclusion agrees with
the result obtained from our foregoing intuitional consideration.

We also would like to mention that under the criterion (\ref{22}) we find
the PWC for a 3-level QHE can be looser than that for a 2-level case. If we
use other criterions, the result may be different. For example, if we use $%
\Delta _{2}$ instead of $\Delta $ in the first inequality of (\ref{22}), we
can not find such 3-level QHE that whose PWC is looser than that for a
2-level case. Actually, for \textbf{Case (I)}, the PWC (\ref{22}) can always
be simplified to inequality (\ref{29}) $T_{h}>T_{l}\left( \Delta ^{h}/\Delta
^{l}\right) \theta $. It can be proved that the coefficient $\left( \Delta
^{h}/\Delta ^{l}\right) \theta $ satisfies%
\begin{equation}
\min \left\{ \frac{\Delta _{1}^{h}}{\Delta _{1}^{l}},\frac{\Delta _{2}^{h}}{%
\Delta _{2}^{l}}\right\} \leqslant \frac{\Delta ^{h}}{\Delta ^{l}}\theta
\leqslant \max \left\{ \frac{\Delta _{1}^{h}}{\Delta _{1}^{l}},\frac{\Delta
_{2}^{h}}{\Delta _{2}^{l}}\right\} .
\end{equation}%
This conclusion is also true for $n$-level QHE. Namely, if the PWC for a $n$%
-level QHE can be expressed as $T_{h}>T_{l}\kappa $, then it can be proved
that the coefficient $\kappa $ satisfies%
\begin{equation}
\min \left\{ \frac{\Delta _{1}^{h}}{\Delta _{1}^{l}},\cdots ,\frac{\Delta
_{n}^{h}}{\Delta _{n}^{l}}\right\} \leqslant \kappa \leqslant \max \left\{ 
\frac{\Delta _{1}^{h}}{\Delta _{1}^{l}},\cdots ,\frac{\Delta _{n}^{h}}{%
\Delta _{n}^{l}}\right\} .
\end{equation}%
Here $\Delta _{i}^{h}$ and $\Delta _{i}^{l}$ are the energy gaps between the 
$i$th and the $\left( i-1\right) $th energy level in \emph{Steps 1 }and 
\emph{Step 3}. Specifically, when all the level spacings change in the same
ratio, i.e., $\Delta _{1}^{h}/\Delta _{1}^{l}=\Delta _{2}^{h}/\Delta
_{2}^{l}=\cdots =\Delta _{n}^{h}/\Delta _{n}^{l}$, no matter how the
spectral structure is, the PWC for such a system has the same form as that
for a 2-level case $T_{h}>T_{l}\left( \Delta ^{h}/\Delta ^{l}\right) $. The
harmonic oscillator and the infinite well potential are two good examples 
\cite{Kieu1,Infinite well}.

\section{Illustration: 3-level atom with dark state}

In the preceding section, we found in proper conditions the PWC for the
3-level QHE can be looser that that for a 2-level QHE (\textit{condition }$I$%
). In this section, we will use a concrete example to demonstrate our
results. We consider a toy model with 3-level system couples to a classical
single-mode external field. The levels are adjusted by the external light
field, which plays a similar role to an ideal piston in classical heat
engine. In this ideal case we need not to consider the work for the
controlling field to adjust the level during the adiabatic process. Just
through this external field the work (either positive or negative) done on
the 3-level system means the change of energy of the 3-level system. Namely,
in our theoretical studies, no extra work is required for the controlling
field since we regarded the field as a part of the external entries .

\bigskip 
\begin{figure}[bp]
\includegraphics[bb=109 56 488 341, width=7 cm, clip]{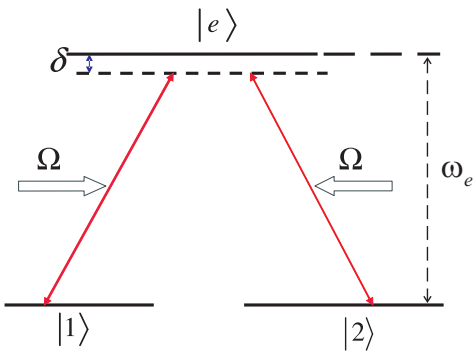}
\caption{(color online) The sketch of a 3-level QHE model. A $\protect%
\lambda $-type 3-level system interacting with a single classical radiation
field. $\left\vert e\right\rangle $ is the exited state, while $\left\vert
1\right\rangle $ and $\left\vert 2\right\rangle $ are the two degenerate
ground states.}
\label{fig.6}
\end{figure}

The model Hamiltonian \cite{Scully2} reads 
\begin{equation}
\widehat{H}=\hbar \delta \left\vert e\right\rangle \left\langle e\right\vert
+\Omega \left( \left\vert e\right\rangle \left\langle 1\right\vert
+\left\vert e\right\rangle \left\langle 2\right\vert +h.c\right) ,
\label{36}
\end{equation}%
where $\delta $\ is the common detuning (see Fig. 10). We have set the
eigenenergy of the two degenerate ground states $\left\vert 1\right\rangle $
and $\left\vert 2\right\rangle $ as zero, and that of the excited state $%
\left\vert e\right\rangle $ as $\hbar \omega _{e}$. $\Omega $ is the complex
Rabi frequency associated with the coupling of the field mode of frequency $%
\omega $ to the atomic transition $\left\vert e\right\rangle \rightarrow
\left\vert 1\right\rangle $ ($\left\vert e\right\rangle \rightarrow
\left\vert 2\right\rangle $). The detuning of $\omega $ is $\delta $,%
\begin{equation}
\omega =\omega _{e}-\delta .  \label{38}
\end{equation}%
We solve the eigen-equation, and obtain\textbf{\ }the eigenvalues%
\begin{eqnarray}
E_{0} &=&0,  \label{39} \\
E_{+} &=&\frac{1}{2}\left( \hbar \delta +\sqrt{\left( \hbar \delta \right)
^{2}+8\left\vert \Omega \right\vert ^{2}}\right) ,  \notag \\
E_{-} &=&\frac{1}{2}\left( \hbar \delta -\sqrt{\left( \hbar \delta \right)
^{2}+8\left\vert \Omega \right\vert ^{2}}\right) .  \notag
\end{eqnarray}%
Then this QHE has two level spacings:%
\begin{eqnarray}
\Delta _{1} &=&E_{0}-E_{-}=\frac{1}{2}\left( \sqrt{\left( \hbar \delta
\right) ^{2}+8\left\vert \Omega \right\vert ^{2}}-\hbar \delta \right) ,
\label{40} \\
\Delta _{2} &=&E_{+}-E_{0}=\frac{1}{2}\left( \sqrt{\left( \hbar \delta
\right) ^{2}+8\left\vert \Omega \right\vert ^{2}}+\hbar \delta \right) . 
\notag
\end{eqnarray}%
The two level spacings are functions of two independent parameters $\delta $%
\ and $\left\vert \Omega \right\vert $. During \emph{Steps 2} the level
spacings change adiabatically from $\Delta ^{h}\left( \Delta _{1}^{h}\right) 
$ to $\Delta ^{l}\left( \Delta _{1}^{l}\right) $ when the two parameters
change slowly from ($\delta ^{h}$, $\left\vert \Omega ^{h}\right\vert $)\ to%
\textbf{\ }($\delta ^{l}$, $\left\vert \Omega ^{l}\right\vert $). According
to the systematical analysis in last section, this model can indeed serve as
an improved 3-level QHE if the two parameters change properly. We plot the
energy levels $\Delta \left( \delta \text{, }\left\vert \Omega \right\vert
\right) $ and $\Delta _{1}\left( \delta \text{, }\left\vert \Omega
\right\vert \right) $ in Fig. 11.

\begin{figure}[tbp]
\includegraphics[bb=293 441 590 689, width=8 cm, clip]{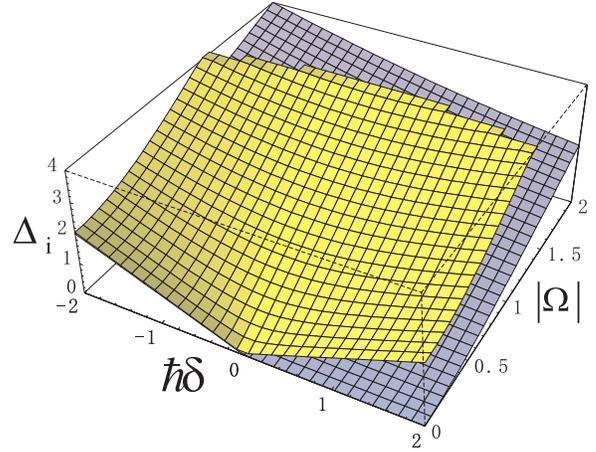}
\caption{(color online) The two level spacings as functions of $\hbar 
\protect\delta $ and $\left\vert \Omega \right\vert $. The upper curved
surface represents $\Delta ^{s}$ while the lower one represents $\Delta
_{1}^{s}$.}
\label{fig.7}
\end{figure}

In a thermodynamic cycle, the level spacing changes adiabatically between $%
\Delta _{i}^{h}$ and $\Delta _{i}^{l}$ during the two adiabatic processes.
For convenience, we define 
\begin{equation}
K^{s}=\sqrt{\left( \hbar \delta ^{s}\right) ^{2}+8\left\vert \Omega
^{s}\right\vert ^{2}}.  \label{40.5}
\end{equation}%
The two constraints of \textbf{Case (I) }in Eq. (\ref{27}) can also be
expressed in terms of $K^{s}$ and $\delta ^{s}$: 
\begin{equation}
\frac{K^{l}-\hbar \delta ^{l}}{K^{h}-\hbar \delta ^{h}}<1,\frac{K^{l}+\hbar
\delta ^{l}}{K^{h}+\hbar \delta ^{h}}<1.  \label{41}
\end{equation}%
Substituting Eq. (\ref{40}) into Eq. (\ref{34}), we get the Solution $I$%
\emph{\ }in terms of $K^{s}$ and $\delta ^{s}$,%
\begin{equation}
\left\{ 
\begin{array}{c}
\delta ^{l}K^{h}>\delta ^{h}K^{l}, \\ 
\delta ^{h}>\delta ^{l}.%
\end{array}%
\right.  \label{41.5}
\end{equation}%
Combine Eqs. (\ref{41}) and (\ref{41.5})\textit{, }we find the inequalities
hold only when $\delta ^{h}>\delta ^{l}>0$. Thus the solution (\ref{41.5})\
can be reduced into a more compact form 
\begin{equation}
\left\{ 
\begin{array}{c}
\delta ^{h}>\delta ^{l}>0, \\ 
\left\vert \Omega ^{l}/\delta ^{l}\right\vert <\left\vert \Omega ^{h}/\delta
^{h}\right\vert .%
\end{array}%
\right.  \label{42}
\end{equation}

\begin{figure}[tbp]
\includegraphics[bb=124 184 436 390, width=8 cm, clip]{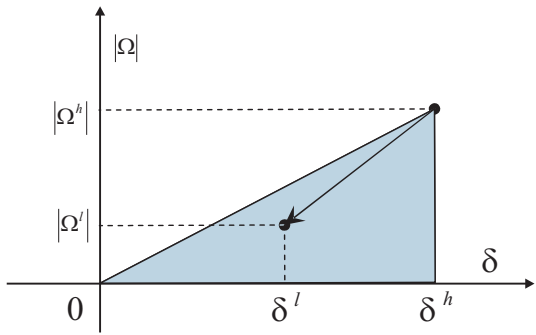}
\caption{(color online) Illustration of Eq. (\protect\ref{42}). In this
solution the QHE can extract positive work while the PWC is looser comparing
with that for 2-level QHE.}
\end{figure}
During \emph{step 2 }of a thermodynamic cycle, we assume the initial level
spacings are $\Delta ^{h}$ and $\Delta _{1}^{h}$ corresponding to two
positive parameters $\delta ^{h}$ and $\left\vert \Omega ^{h}\right\vert $.
Then the level spacings change adiabatically to $\Delta ^{l}$ and $\Delta
_{1}^{l}$ when the two parameters change slowly to ($\delta ^{l}$, $%
\left\vert \Omega ^{l}\right\vert $). The above result (\ref{42}) shows
that, only when the final point ($\delta ^{l}$, $\left\vert \Omega
^{l}\right\vert $) locates in the shaded triangle in Fig. 12 can we make the
PWC looser compared to \textit{condition }$I\ $for a 2-level case.

\begin{figure}[tbp]
\includegraphics[bb=75 55 369 256, width=8 cm, clip]{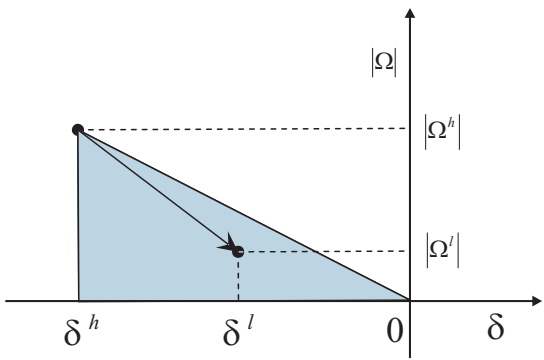}
\caption{(color online) Illustration of Eq. (\protect\ref{43}). The physical
meaning is similar to that of Fig. 12. But this is not an ideal solution.}
\label{fig.9}
\end{figure}

Similarly, we substitute Eq. (\ref{40}) into Eq. (\ref{35}) and then obtain
Solution $II$\ in terms of $K^{s}$ \ and $\delta ^{s}$.%
\begin{equation}
\left\{ 
\begin{array}{c}
\delta ^{h}<\delta ^{l}<0, \\ 
\left\vert \Omega ^{l}/\delta ^{l}\right\vert <\left\vert \Omega ^{h}/\delta
^{h}\right\vert .%
\end{array}%
\right.  \label{43}
\end{equation}%
The physical meaning is almost the same as that of Fig. 12. For the initial
parameters $\delta ^{h}$ and $\left\vert \Omega ^{h}\right\vert $, only
when\ the final point ($\delta ^{l}$, $\left\vert \Omega ^{l}\right\vert $)
locates in the shaded triangle in Fig. 13 can the first inequality of (\ref%
{22}) be satisfied, but the second is still not. Thus Solution $II$ is not a
desired solution.

Before concluding this section, we would like to address a crucial problems
in the physical implementation of the above \textquotedblleft dark state
model". It is obvious that we do work on the system when varying the
parameters $\delta $ and $\left\vert \Omega \right\vert $ in step 4. But how
is the work extracted out of the system can not be imagined directly. Here,
what we discussed in this section is only a toy mode to illustrate the basic
spirit of physics in the quantum thermodynamic cycles. The work done by the
system in step 2 can be indirectly understood as the consequence of the
decrease of the system energy, which can be explicitly calculated from our
foregoing analysis (see Eq. (\ref{6})) in section II.

\section{Conclusion and remarks}

We have quantum-mechanically formulated the work done and the heat
transferred in the thermodynamic processes in association with the
microscopic quantum transitions. A class of QHE were universally proposed by
using a multi-level quantum system as the working substance and by deriving
the thermodynamic adiabatic process from the quantum adiabatic process. We
classified a 3-level QHE based on the changes of the level spacings, and
found when the parameters (and thus the level spacings) change properly, a
3-level QHE can better the work extraction compared to a 2-level case.

Before concluding this paper we would like to point out that the
two-parameter QHE proposed in the last section is only a toy model and one
can not over estimate it. We have to say that the Hamiltonian for such an
atomic system in the laboratory frame of reference is time-dependent and
changes fast. Therefore one can roughly regarded it as something relevant to
the system energy. It is still an open question to find a multi-level system
with level spacings changing independently in practice.

We also remark that the discussion about QHE is essentially semi-classical
because we quantize neither the heat bath nor the controllable external
fields. To build a totally-quantum theory for QHE, one need to use the
generalized master equations with and without memories. There are some
interesting results in Refs. \cite{Bender1,Bender2}. How to develop our
present studies within this theoretical framework is to be considered in our
forthcoming investigations.

\noindent \textbf{Acknowledgement:} This work is supported by the NSFC with
grant Nos. 90203018, 10474104 and 10447133. It is also funded by the
National Fundamental Research Program of China with Nos. 2001CB309310 and
2005CB724508. We thank T. D. Kieu for helpful discussion.


\begin{thebibliography}{99}
\bibitem{book} D. P. Sheehan (Ed), \textit{Quantum limits to the second law:
first international Conference}, (Melville, New York, 2002).

\bibitem{Scully-science} M. O. Scully, M. S. Zubairy, G. S. Agarwal, H.
Walther, \textit{Science} \textbf{299}, 862 (2003).

\bibitem{Scully-book} M. O. Scully, \textit{Quantum limits to the second
law: first international Conference}, edited by D.P.Sheehan (2002).

\bibitem{Zubairy-book} M. S. Zubairy, \textit{Quantum limits to the second
law: first international Conference}, edited by D.P.Sheehan (2002).

\bibitem{Kieu-PRL} T. D. Kieu, \textit{Phys. Rev. Lett.} \textbf{93}, 140403
(2004).

\bibitem{Kieu1} T. D. Kieu, \textit{arXiv:quantum-ph/0311157 v5} 22 Aug 2005.

\bibitem{Kosloff1} T. Feldmann and R. Kosloff, \textit{Phys. Rev. E} \textbf{%
61}, 4774 (2000).

\bibitem{Infinite well} Similar to the harmonic oscillator, it is easy to
prove that the PWC for the infinite square well model is $T_{h}>T_{l}\left(
L_{l}/L_{h}\right) ^{2}$, where $L_{l}$ and $L_{h}$ are the widths of the
well in two steps.

\bibitem{Arnaud} J. Arnaud, L. Chusseau, F. Philippe, \textit{%
arXiv:quantum-ph/0211072 v2} 2 Jun 2003.

\bibitem{He} J. He, J. Chen, B. Hua, \textit{Phys. Rev. E }\textbf{65},
036145 (2002).

\bibitem{Kosloff2} R. Kosloff, T. Feldmann, \textit{Phys. Rev. E }\textbf{65}%
, 055102(R) (2002).

\bibitem{Kosloff3} E. Geva and R. Kossloff, \textit{J. Chem. Phys.} \textbf{%
96}, 3054 (1992).

\bibitem{yukawa} H. Yukawa (Ed), \textit{Quantum Mechanics} Vol I, 2nd Ed,
(in Japanese), (Yan-Bo Bookshop, Tokyo, 1978).

\bibitem{adiabatic condition} C. P. Sun, \textit{J. Phys. A} \textbf{21},
1595 (1988), \textit{Phys. Rev. D} \textbf{41}, 1318 (1990).

\bibitem{Bohr} L. I. Schiff, \textit{Quantum Mechanics, }3rd ed.\textit{, }%
(McGRAW-HILL, Inc., New York, 1968).

\bibitem{Scully2} M. O. Scully and M. S. Zubairy, \textit{Quantum Optics }%
(Cambridge University Press, 1997).

\bibitem{Bender1} C. M. Bender, D. C. Brody, B. K. Meister, \textit{Proc. R.
Soc. Lond. A} \textbf{458},1519 (2002).

\bibitem{Bender2} C. M. Bender, D. C. Brody, B. K. Meister, \textit{J. Phys.
A} \textbf{33}, 4427 (2000).
\end{thebibliography}
\end{document}